\documentclass[twocolumn,12pt,cites,graphicx]{article}
\usepackage{epsfig}
\usepackage{graphicx}
\usepackage[square,numbers,sort&compress]{natbib}
\usepackage{subfigure}

\newcommand{\ie}{\emph{i.e.},~}

\title{Polyelectrolytes in electric fields: Measuring the dynamical effective
charge and effective friction}

\author{
Kai Grass\thanks{Frankfurt Institute for Advanced Studies, Goethe University,
Ruth-Moufang-Strasse 1, D-60438 Frankfurt am Main, Germany.
E-mail: grass@fias.uni-frankfurt.de} ~and Christian
Holm$^{*,}$\thanks{Max-Planck Institute for Polymer Research, Ackermannweg 10, D-55128 Mainz, Germany.
E-mail: c.holm@fias.uni-frankfurt.de} }

\date{Received XXXXth Month, 200X\\Accepted XXXXth Month, 200X\\DOI: 10.1039/}

\begin{document}

\maketitle
\renewcommand{\thefootnote}{\fnsymbol{footnote}}

\noindent We use a coarse-grained molecular dynamics model to study
the electrophoretic behaviour of flexible polyelectrolyte chains. We
first characterize the static properties of the model with respect to
the chain length, the polyelectrolyte concentration, additional salt
and the influence of an applied external field. Next we investigate
the dynamic behaviour in the oligomer range and compare to data
obtained by two different experimental methods, namely capillary
electrophoresis and PFG-NMR. We find excellent agreement of
experiments and simulations when hydrodynamic interactions are
accounted for in the simulations.  We then present novel estimators
for the dynamical effective charge during free solution
electrophoresis and compare them to static estimators. We 
find complete agreement between the static and the dynamic
estimators. We further evaluate the scaling behaviour of the effective
friction of the polyelectrolyte-counterion complex with the
surrounding fluid. We identify a hydrodynamic screening length beyond
which the friction during electrophoresis is linear depending on the
chain length resulting in a constant mobility for long polyelectrolyte
chains. Our results show a convincing agreement with experimental data
and demonstrate that it is possible to model dynamic behaviour of
polyelectrolytes using coarse grained models, provided they include
the effects of hydrodynamical interactions.

\section{Introduction}

In order to be able to improve the processes involved in current
electrophoretic separation methods it is a prerequisite to gain a
thorough understanding of the behaviour of polyelectrolytes in an
externally applied electric field. Electrophoretic methods are widely
used to study polyelectrolytes (PEs) such as proteins, DNA and
synthetic polymers~\cite{righetti96a,cottet05a,dolnik06a}. Several
theories~\cite{barrat96a,muthukumar96a,volkel95b,mohanty99a} have been
developed to describe PE electrophoresis and successfully described
qualitatively the experimentally observed behaviour of various PEs
under bulk conditions. In a recent publication~\cite{grass08a}, we
showed how coarse-grained molecular dynamics simulations can extend
this theoretical understanding on a more microscopic level. In this
article, we will extend this work by rigorously studying additional
effects on the electrophoretic behavior such as polyelectrolyte
concentration, concentration of added salt, and variations of the
externally applied field.

We start to characterize the polyelectrolyte's static properties on
the externally applied field. We then continue to investigate the
dynamic properties, putting emphasis on the hydrodynamic interactions
between the polyelectrolyte and the surrounding solvent. Our results
on the PE diffusion coefficient and the mobility are compared to two
experimental data sets and yield very good agreement. Since the
mobility of the PE can be expressed as the quotient of an effective
charge and an effective friction term, we concentrate on finding a
reasonable operational definition of an effective charge. This would
then yield a measure of the effective friction, and we believe that a
direct measurement of the effective charge will help to fully
understand the electrophoretic process. We also would like to clarify
if there are differences in the obtained values of the effective
charge, when purely dynamical or static estimators are used, as has
been found for charged colloidal suspensions~\cite{palberg04a}.  We
compare five different ways to estimate the effective charge of the
polyelectrolyte-counterion complex, and use the effective charge to
characterize the effective friction, thereby identifying the role of
counterions on hydrodynamic shielding.

The paper is organised as follows: in Section~\ref{sec:model} we
present our simulation model and briefly describe the different
systems investigated. The model is characterised in
Section~\ref{sec:staticprops} by analyzing the dependence of static
system properties on the the system parameters such as chain lengths
and concentration, effect of added salt, and the influence of the
external electric field applied during electrophoresis. We then
determine two relevant transport coefficients, the diffusion and the
electrophoretic mobility, and compare the results to recent
experimental data (Section~\ref{sec:transportcoeff}). To understand
the observed results, we introduce and compare several estimators for
the effective charge of the polyelectrolyte-counterion-complex that is
created in solution (Section~\ref{sec:chargeestimators}). In
Section~\ref{sec:effectivefriction}, we use the charge estimate to
calculate the effective friction of the complex during free solution
electrophoresis. We identify a linear increase of the friction with
the chain length beyond a screening length. We conclude the paper by
interpreting the observed results and show how this helps to
understand the underlying processes in free solution electrophoresis.

\section{Model}\label{sec:model}

We employ molecular dynamics (MD) simulations using the ESPResSo
package~\cite{limbach06a} to study the behaviour of linear PEs of different
lengths. The PEs are modelled by a totally flexible bead-spring model in a set of
spheres that represent the $N$ individual monomers which are connected to each
other by finitely extensible nonlinear elastic (FENE) bonds~\cite{soddeman01a}
\[
U_\mathrm{FENE}(r) = \frac{1}{2} k R^2 \ln \left( 1 - \left( \frac{r}{R} \right)^2
\right),
\]
with stiffness $k = 30 \epsilon_0$, and maximum extension $R = 1.5
\sigma_0$, and $r$ being the distance between the interacting monomers.
Additionally, a truncated Lennard-Jones or WCA potential~\cite{weeks71a}
\[
U_\mathrm{LJ}(r<r_\mathrm{c}) = \epsilon_0 \left( \left(
\frac{\sigma_0}{r}\right)^{12} - \left(
\frac{\sigma_0}{r}\right)^6 + \frac{1}{4} \right),
\]
is used for excluded volume interactions. A cutoff value of $r_c =
\sqrt[6]{2}\sigma_0$ ensures a purely repulsive potential. All dissociated counterions and
additional salt ions are modelled by appropriately charged spheres using the same WCA potential.

Here, $\epsilon_0$ and $\sigma_0$ define the energy and length scale of the
simulations. We use $\epsilon_0 = k_\mathrm{B} T$, i.e.~the energy of the system is
expressed in terms of the thermal energy. The length scale $\sigma_0$ defines the
size of the monomers. In this model, the average bond length between the
successive monomers in a PE chain is $0.91 \sigma_0$, whereas the real distance
between two monomers of a fully sulfonated polystyrene chain is approximately 2.5
\AA. Therefore, $1 \sigma_0$ is mapped to  $2.5 \mathrm{\AA}/0.91 = 2.75
\mathrm{\AA}$ in reality. Different polyelectrolytes can be mapped by changing
$\sigma_0$. Unless mentioned otherwise, all observables are expressed in reduced
simulation units, and we will not use $\sigma_0$ and $\epsilon_0$ explicitly from
now on.

All chain monomers carry a negative electric charge $q = -1 e_0$,
where $e_0$ is the elementary charge. For charge neutrality, $N$
monovalent counterions of charge $+1 e_0$ are added. Where mentioned,
additional monovalent salt is added to the simulation. Full
electrostatic interactions are calculated with the P3M algorithm using
the implementation of Reference ~\cite{deserno98a}. The Bjerrum length
$l_\mathrm{B} = e_0^2 / \left( 4 \pi \epsilon_0 \epsilon_\mathrm{r}
  k_\mathrm{B} T \right)= 2.58$ in simulation units corresponds to 7.1
\AA, the Bjerrum length in water at room temperature. This means that
the effect of the surrounding water is modelled implicitly by simply
using the dielectric properties of water, having a relative dielectric
constant of $\epsilon_r \approx 80$.

The simulations are carried out under periodic boundary conditions in a
cubic simulation box. The size $L$ of the box is varied to realize a
constant monomer concentration $c_\mathrm{PE}$ independent of chain length as specified.

We compare two types of MD simulations. In the first one, we use simple
Langevin equations of motions with a velocity dependent dissipative and a
random term in addition to the interparticle forces. Together, both additional
terms implicitly model the effects of a solvent surrounding the particles: the
dissipative force, $\mathbf{F_\mathrm{D}} = -\Gamma_0 \mathbf{v}$, provides
local friction and the non-correlated zero-mean Gaussian random forces,
$\mathbf{F_\mathrm{R}} = -\mathbf{\xi}(t)$, mimic thermal kicks (Brownian
motion). In order to fulfill the fluctuation-dissipation theorem, dissipative
and random force have to be coupled together:
$\langle \xi_i(t) \cdot \xi_j(t') \rangle = 6 \Gamma_0 k_\mathrm{B} T
\delta_{ij} \delta(t-t')$. This approach only offers local particle-fluid
interactions, but destroys long-range hydrodynamic interactions (HI).

The second set of simulations include hydrodynamics using a Lattice Boltzmann
(LB) algorithm~\cite{mcnamara88a} that is interacting with the MD
simulations via a frictionally coupling introduced by Ahlrichs et al.~\cite{ahlrichs99a}.
The mesoscopic LB fluid is described by a velocity field generated by
discrete momentum distributions on a spatial grid rather than explicit fluid
particles. We use an implementation of the D3Q18 model with a kinematic
viscosity $\nu = 3.0$, and a fluid density $\rho = 0.864$. The resulting fluid has
a dynamic viscosity $\eta = \rho \nu = 2.592$.
The space is discretised by a grid with spacing $a = 1.0$. As in the
Langevin approach, the particle-fluid interaction is realised by a dissipative
force depending on the difference between the particle velocity $\mathbf{v}$
and the fluid velocity at the particle position $\mathbf{u}$:
$\mathbf{F_\mathrm{R}} = -\Gamma_\mathrm{bare} (\mathbf{v}-\mathbf{u})$. Here,
the coupling constant is $\Gamma_\mathrm{bare} = 20.0$. Additional random
fluctuations for particles and fluid act as a thermostat. The interaction
between particles and fluid conserve total momentum, and this algorithm has
been shown to yield correct long-range hydrodynamic interaction
between individual particles\cite{ahlrichs99a}.

By comparing both types of MD simulations, we can characterize the impact of HI
on the dynamics of the system. The individual approaches are mapped by setting the
Langevin friction parameter $\Gamma_0 = 15.34$ to match the single particle
mobility obtained with the mesoscopic fluid.

The simulations are carried out with a MD time step $\tau_\mathrm{MD}
= 0.01$ and LB time step $\tau_\mathrm{LB} = 0.05$. After an
equilibration time of $10^6$ steps, $10^7$ steps are used for
generating the data. The time-series are analyzed using
auto-correlation functions to estimate the statistical errors as
detailed in Reference~\cite{wolff04a}. Error bars of the order of the
symbol size or smaller are omitted in the figures. Up to ten
independent simulations are carried out for each data point, taking
between one day and three weeks on a single standard CPU depending on
the chain length $N$ and the monomer concentration $c_\mathrm{PE}$.

\section{Static properties}\label{sec:staticprops}

In this section, the proposed coarse-grained model is used to quantify the
static properties of the polyelectrolyte chain and the surrounding counterions. The
influence of the individual concentrations of the monomers $c_\mathrm{PE}$ and
the counterions $c_\mathrm{CI}$ as well as the dependence on the strength of the
external electric field $E$ are investigated.

\subsection{Chain scaling}\label{sec:chainscaling} 	


The polyelectrolyte chain conformations can be characterised by the average
end-to-end distance
\[
R_\mathrm{e}^2 = \left< \left( \vec{r}_{1} - \vec{r}_{N} \right)^2
\right>,
\]
the radius of gyration
\[
R_\mathrm{g}^2 = \left< \frac{1}{N} \sum_{i}\left( \vec{r}_{i}-\vec{r}_{\mathrm{cm}}
\right)^2 \right>,
\]
and by the hydrodynamic radius 
\[
  \left< \frac{1}{R_\mathrm{h}} \right> = \frac{1}{N} \sum_{i \neq j}
  \left< \frac{1}{\| \vec{r}_{\mathrm{i}}-\vec{r}_{\mathrm{j}}\|} \right>,
\]
Here, $N$ is the number of chain monomers, $\vec{r}_{i}$ the position
of the $i$-th monomer, and $\vec{r}_{\mathrm{cm}}$ the center of mass of the
polyelectrolyte chain. The angular brackets $\langle \ldots \rangle$ indicate
an ensemble average. All three quantities are expected to exhibit a power law
scaling $R_\mathrm{e,g,h} \sim \left(N-1\right)^\nu$, where the scaling exponent
$\nu$ depends on the system. For an uncharged polymer with ideal
chain behaviour $\nu \approx 0.588$ (Flory exponent)~\cite{moore78a},
whereas for a fully charged polyelectrolyte without electrostatic screening (no
counterions), we expect to find $\nu = 1$.

\begin{figure}[ht]
\begin{center}
  \includegraphics[width=1.0\columnwidth]{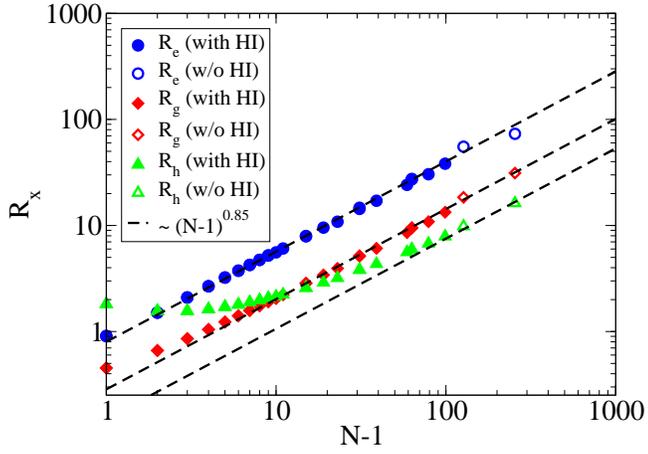}
  \caption{Both the end-to-end distance
  $R_\mathrm{e}$ and the radius of gyration $R_\mathrm{g}$ at vanishing external field
  for a monomer concentration of
  $c_\mathrm{PE} = c_\mathrm{CI} = 10$ mM exhibit an effective scaling exponent $\nu =
  0.85$. The hydrodynamic radius $R_\mathrm{h}$ shows a different behaviour for short chains
  and only slowly reaches the asymptotic scaling.
  The static chain properties are not influenced by hydrodynamic
  interactions (HI) and fully agree with the ones obtained without HI.}
  \label{fig:rxscaling}
\end{center}
\end{figure}

Figure~\ref{fig:rxscaling} shows $R_\mathrm{e,g,h}$ for
polyelectrolyte chains of different length in the presence of the
neutralizing counterions at a concentration for monomers and
counterions of $c_\mathrm{PE} = c_\mathrm{CI} = 10$ mM and in the
absence of an external field. The obtained effective scaling exponent
$\nu \approx 0.85$ for $R_\mathrm{e}$ and $R_\mathrm{g}$ for chains up
to $N=256$ indicates the counterion influence on electrostatic
screening and on the chain conformations. The hydrodynamic radius
$R_\mathrm{h}$ exhibits a very slow asymptotic behaviour which leads
to a lower apparent scaling exponent for short and intermediate
chains. In Section~\ref{sec:diffusion}, we will demonstrate, however,
that this is in perfect agreement with the measured diffusion
coefficients.  Figure~\ref{fig:rxscaling} also examplifies that there
are no influences of the hydrodynamic interactions on all static chain
properties, as it should be.


If the polyelectrolyte chain is subject to a strong external electric field
$E$, then, depending on the magnitude of $E$, conformational changes can be
induced~\cite{netz03c}.

\begin{figure}[ht]
\begin{center}
  \subfigure[]{
    \includegraphics[width=0.95\columnwidth]{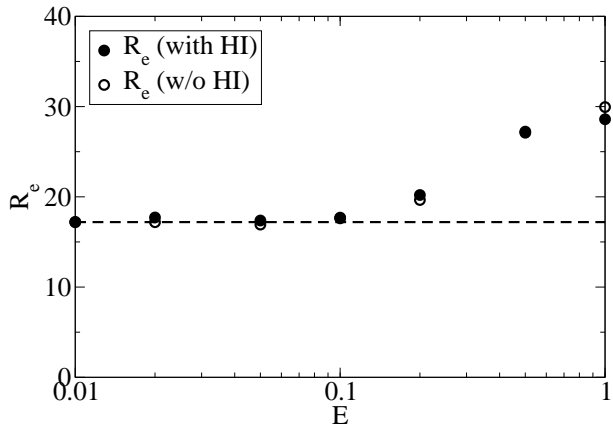}
    \label{fig:reE}
  }
  \subfigure[]{
    \includegraphics[width=0.95\columnwidth]{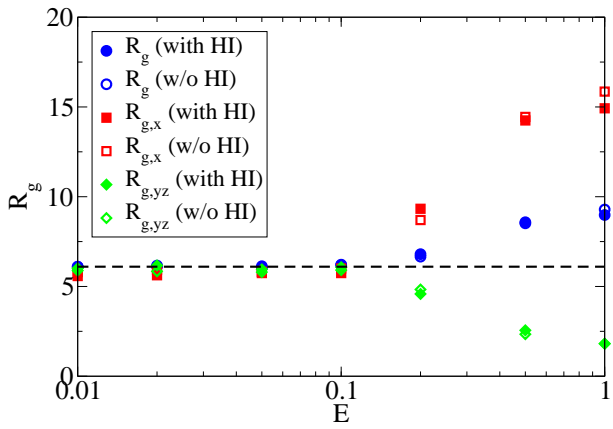}
    \label{fig:rgE}
  }
  \caption{(a) The end-to-end distance $R_\mathrm{e}$ and (b) the radius of
  gyration radius $R_\mathrm{g}$ for a chain of $N=40$ monomers differ from the
  value at zero external field $E$ (dashed line), if a critical
  value is reached. Beyond this threshold, the polyelectrolyte chain is extended
  (increased $R_\mathrm{e}$ and $R_\mathrm{g}$) and starts to align with the field
  directed in x-direction. This effect is independent of hydrodynamic interactions (HI).}
  \label{fig:rxE}
\end{center}
\end{figure}

In Figure~\ref{fig:rxE}, we show that the model polyelectrolyte of length
$N=40$ experiences conformational changes if the external electric field is
$E_\mathrm{crit} \approx 0.2$ or larger. According to~\cite{netz03b,netz03c},
$E_\mathrm{crit}$ depends on the strength of the electrostatic coupling and
on the length $N$. The approximation of Netz $E_\mathrm{crit} \approx
\sqrt{l_\mathrm{B} / N} \approx 0.25$ yields a value comparable to the one
found in our simulations.

The increased end-to-end distance (Figure~\ref{fig:reE}) indicates that the
polyelectrolyte chain conformation is extended. This can also be seen by
looking at $R_\mathrm{g}$ in Figure~\ref{fig:rgE}. Additionally, we introduce two
new observables
\[
R_\mathrm{g,x}^2 = \left<
 \frac{3}{N} \sum_{i}\left( {r_{i\mathrm{,x}}}-{r_\mathrm{cm,x}}
\right)^2 \right>
\]
and
\[
R_\mathrm{g,yz}^2 = \left<
\frac{3}{2N} \sum_{i,\alpha=\mathrm{y,z}}\left(
{r_{i\mathrm{,}\alpha}}-{r_{\mathrm{cm,}\alpha}} \right)^2 \right>,
\]
where $r_{i,\alpha}$ is the position component in $\alpha$-direction of the
 $i$-th particle. For an isotropic chain, $R_\mathrm{g} = R_\mathrm{g,x} = R_\mathrm{g,yz}$,
which is shown in Figure~\ref{fig:rgE} for $E \leq 0.1$. For electric fields
beyond the threshold, an increase of $R_\mathrm{g,x}$ and a decrease of
$R_\mathrm{g,yz}$ is observed. This can be understood as an extension of the
polyelectrolyte conformation in x-direction (into the direction of the external
field) and a compression perpendicular to it. This alignment in high electric
fields has also been studied in Reference~\citealp{schlagberger05a}. There it
was shown, that for even higher fields and stiff polymers, the phase of alignment along the
electric field is followed by an alignment perpendicular to it. This transition
has also been recently observed by~\cite{frank08a}. Figure~\ref{fig:rxE} also
shows that this effect is independent of long range hydrodynamic
interactions, which is consistent with the initial observations by Netz that
were obtained without the inclusion of hydrodynamic
interactions~\cite{netz03b,netz03c}.

For the purpose of this paper, we note that for electrical fields of $E = 0.1$
or lower, no conformational change and no orientational order is induced.

\begin{figure}[ht]
\begin{center}
  \includegraphics[width=1.0\columnwidth]{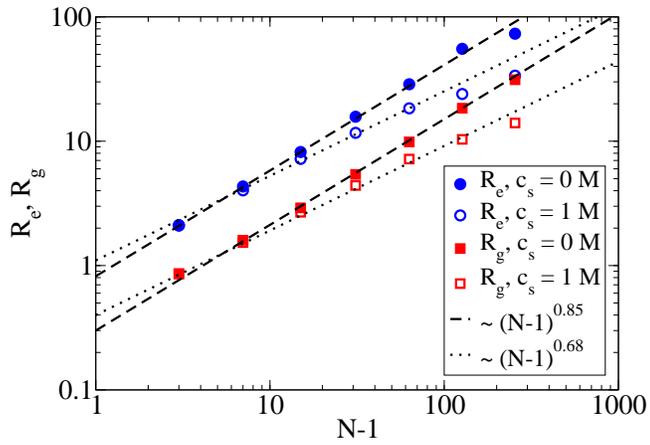}
  \caption{The end-to-end distance
  $R_\mathrm{e}$ and the radius of gyration $R_\mathrm{g}$ at zero external
  field and for a monomer concentration $c_\mathrm{PE} = 10$ mM and no added salt
  ($c_\mathrm{s} = 0$ M) exhibit an effective scaling exponent $\nu = 0.85$ (dashed
  line). With $c_\mathrm{s} = 1$ M monovalent salt added, the scaling exponent drops down to
  $\nu = 0.68$ (dotted line).}
  \label{fig:rx-salt}
\end{center}
\end{figure}

Adding salt to a polyelectrolyte solution screens
electrostatic interactions on a length scale known as the Debye length
$l_\mathrm{D}$. For monovalent salt, $l_\mathrm{D}$ is inversely
proportional to the square root of the added salt concentration:
\[l_\mathrm{D}^{-1} = \sqrt{4 \pi l_\mathrm{B} \left( 2 c_\mathrm{s} + c_\mathrm{CI}
\right) N_\mathrm{A}},\]
where $c_\mathrm{s}$ is the concentration of the monovalent
salt, $c_\mathrm{CI}$ is the concentration of the polyelectrolyte's counterions and
$N_\mathrm{A}$ is the Avogadro number. In Figure~\ref{fig:rx-salt}, the screening
effect of added salt can be seen. We compare the scaling of $R_\mathrm{e}$ and
$R_\mathrm{g}$ for the salt-free case ($l_\mathrm{D} \approx 16.7$) to a solution
with $c_\mathrm{s} = 1$ M of added salt ($l_\mathrm{D} \approx 1.2$). The additional
salt screens the electrostatic interactions between the polyelectrolyte
monomers and reduces the scaling coefficient to $\nu = 0.68$, which is close to
the scaling for an uncharged polymer with the Flory exponent. It remains to be
pointed out, that the measured scaling exponent is only an effective value
for the length range investigated. For chains with a blob diameter 
much larger than the finite electrostatic screening length $l_\mathrm{D}$, the scaling
is expected to be equal to the scaling of an uncharged polymer, $\nu = 0.588$. 

\begin{figure}[ht]
\begin{center}
  \includegraphics[width=1.0\columnwidth]{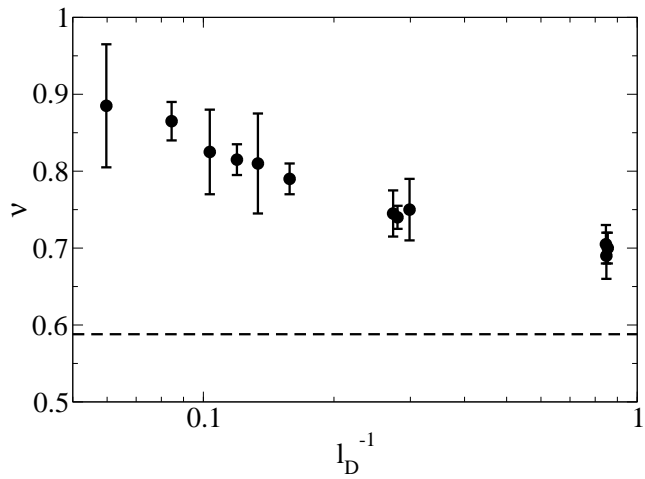}
  \caption[]{The effective scaling factor $\nu$ shows strong dependence of the
  Debye length. With out added salt (i.e.~large Debye lengths) $\nu$ is close to 1, whereas
  for high salt concentrations (i.e.~small Debye lengths) approaches the
  Flory number, the value of an uncharged polymer.}
  \label{fig:nu-salt}
\end{center}
\end{figure}

By adding salt, the polyelectrolyte conformations become less
extended. The effective scaling exponent $\nu$ depends on the inverse
Debye length in the system. We illustrate this in
Figure~\ref{fig:nu-salt}: for no additional salt (small inverse Debye
lengths), the observed scaling exponent is close to 1 as it is
expected for an unscreened polyelectrolyte chain. As we add additional
salt and thereby increase the inverse Debye length, the scaling
exponent decreases and approaches the Flory number. The additional
salt screens the electrostatic interactions along the polyelectrolyte
chain which starts to assume configurations close to those of an
uncharged polymer in ideal solvent.

We note that adding monovalent salt to the solution, decreases the spatial
extension of the polyelectrolyte chain and makes the conformations more compact.

\subsection{Counterions}\label{sec:counterions}


After having investigated the chain conformations in some detail, we
now take a look at the counterion distribution around the
polyelectrolyte chains. Strongly charged polyelectrolytes attract some
of the released counterions and effectively reduce their line charge
density. This phenomenon was described by Manning and Oosawa under the
term counterion condensation~\cite{manning69a,oosawa71a}. This topic
has been discussed from varying viewpoints
(see~\cite{manning98a,stigter95a} and the references therein).

According to Manning's theory, the distribution of counterions around highly charged
rodlike polyelectrolytes can be described in terms of the Manning parameter $\xi
= l_\mathrm{B}/b,$ where $l_\mathrm{B} = 2.58$ is the Bjerrum length
and $b = 0.91$ is distance between charges along the backbone of the
polyelectrolyte (inverse line charge density), thus $\xi=2.84$.

Here, we investigate highly charged polyelectrolytes with $\xi > 1$, for which a
finite number of counterions is always found in close vicinity of the
polyelectrolyte chain, thus reducing the effective charge of the created
polyelectrolyte-counterion complex. The simple counter ion condensation theory predicts the fraction of those
condensed counterions to be $f_\mathrm{CI} = 1 - 1/\xi$, but without specifying
the actual distance to the chain in which those counterions are to be found.
Following the prediction, the total number is
\begin{equation}
  \label{eq:manningnci}
  N_\mathrm{CI} = \left( 1 - 1/\xi \right) N,
\end{equation}
where $N$ is the length of the polyelectrolyte. This leads to a predicted
effective charge $Q_\mathrm{eff}(N) = 1/\xi N$.

In Figure~\ref{fig:nci}, we compare this prediction to the average
number of counterions that we find within a distance of $2 \sigma$ to the
polyelectrolyte chain. We will present more elaborate estimators based on Poisson-Boltzman theory in
Section~\ref{sec:chargeestimators}. For long polyelectrolyte chains ($N >
100$), the measured number of condensed counterions approaches the predicted Manning
value. The distribution of counterions around the chain is not influenced by
hydrodynamic interactions, as it should be.


An external electric field not only couples to the polyelectrolyte monomers,
but also acts on the oppositely charged counterions. Strong electric fields are
known to reduce the number of condensed counterions in the vicinity of the
chain~\cite{netz03b,netz03c}. In Figure~\ref{fig:nciE}, the number of
counterions within $2 \sigma_0$ around a polyelectrolyte chain with $N=40$
monomers is determined for different values of the applied electric field.
Beyond a critical threshold of $E=0.1$, counterions are stripped away from the
polyelectrolyte chain. The onset of this effect coincides with the
observed extension and alignment of the chain (Figure~\ref{fig:rxE}).

Again, we note that for electrical fields of $E = 0.1$
or lower, no change to the counterion distribution around the polyelectrolyte
chain is found.

\section{Transport coefficients}\label{sec:transportcoeff}

In this section, we will determine the transport coefficients for 
flexible polyelectrolytes. We will briefly review the results from an
earlier study combining different experimental data sets on
polystyrene sulfonate (PSS) with simulation
results~\cite{grass08a}. We augment the study by an additional
investigation about the dynamical and static effective charge and friction, respectively,
to expand the understanding of the microscopic processes.

\subsection{Diffusion}\label{sec:diffusion}
	
The diffusion coefficient $D$ characterizes the thermal motion of the
polyelectrolyte. It can be obtained from the simulation trajectory of the
polyelectrolyte chain by recording the center of mass
mean-square displacement versus time, and subsequently measuring the
corresponding slope according to 
\begin{equation}
	\label{eq:diffmsd}
	D = \frac{\left<\left[
	\vec{r}_{\mathrm{cm}}(t)-\vec{r}_{\mathrm{cm}}(0)\right]^2 \right>}{6 t},
\end{equation}
where $\vec{r}_{\mathrm{cm}}$ is the position of the center of mass, and $t$ is
the time. The angular brackets $\langle \ldots \rangle$ indicate the averaging over
many configurations.

Alternatively, the diffusion coefficient $D$ can be obtained from the
integration of the velocity auto-correlation function of the center of mass
\begin{equation}
    \label{eq:diffvac}
	D = \frac{1}{3} \int_0^\infty \left< \vec{v}_{\mathrm{cm}}(t) \cdot
	\vec{v}_{\mathrm{cm}}(0) \right> dt.
\end{equation}
Here, $\vec{v}_{\mathrm{cm}}$ is the center of mass velocity of the
polyelectrolyte at a given time. Again, the angular brackets $\langle \ldots \rangle$ indicate
the averaging over many configurations.


The accuracy of both methods depends on the number of statistically independent
data samples. In Figure~\ref{fig:diffmethods}, we present sample graphs to
determine the diffusion of a polyelectrolyte chain of length $N=32$ using
Equations~\ref{eq:diffmsd} and \ref{eq:diffvac}. Since simulations with
hydrodynamic interactions are computationally very demanding, the achievable
accuracy is limited. The errors are determined from the statistical
fluctuations of the data and the uncertainty in the fit parameter.

In Figure~\ref{fig:diffmethodsa}, the diffusion is obtained from a fit
to the linear part of the mean square displacement. This yields a
diffusion coefficient of $D = 0.0045 \pm 0.0002$ in simulation
units. To calculate the integral in Equation~\ref{eq:diffvac}, a fit
to the slowly decaying long-time tail of the center of mass' velocity
auto-correlation function has to be obtained as shown in
Figure~\ref{fig:diffmethodsb}. Here, the theoretical predicted
functions are used to match the long-time tail: without hydrodynamic
interactions, an exponential decay of the velocity correlations is
expected, whereas with hydrodynamic interactions the correlation
function decays with $t^{-3/2}$. The figure shows the more interesting
case with hydrodynamic interactions. The combined results of
simulation data and long-time fit are integrated and a diffusion
coefficient of $D = 0.0041 \pm 0.0005$ is obtained, which is in
agreement with the corresponding value obtained from the mean square
displacement.

Both methods are strictly equivalent for classical systems, but for the remainder
of the paper, the integral method is used to obtain the diffusion coefficient
as one can use a similar formulation to obtain the mobility of the
polyelectrolyte (see Equation ~\ref{eq:mobilitygreenkubo}), and thus can
determine both quantities without additional computational effort.
Specifically, both auto-correlation functions have to be determined accurately
on the interval $t = [0,100]$, whereas the root mean square displacement has
to be determined on the interval $t= [100,100000]$, as shown in
Figure~\ref{fig:diffmethods}.

The diffusion of polymers in the presence or absence of hydrodynamic
interactions is very well studied from a theoretical point of view. In
general, the following Einstein equation is valid
\[
  D = \frac{k_B T}{\Gamma},
\]
where $\Gamma$ is the so-called friction coefficient of the studied object,
$k_B$ is the Boltzmann factor and $T$ is the temperature.

Without hydrodynamic interactions, one expects Rouse behaviour, i.e.~ a
friction coefficient $\Gamma$ that linearly depends on the change length $N$:
\[
D = \frac{k_B T}{\Gamma_0 N}.
\]
Here, $\Gamma_0$ is the friction coefficient of a single monomer of the polymer
chain.

With hydrodynamic interactions, Zimm behaviour is expected. The scaling of $D$
with the chain length is no longer proportional to $N^{-1}$, but can be
described by the Kirkwood-Zimm theory~\cite{kirkwood48a,zimm56a}. Within this theory, the
diffusion coefficient is expected to be
\begin{equation}
	\label{eq:kirkwooddiffusion}
	D = \frac{D_0}{N} + \frac{k_b T}{6 \pi \eta R_\mathrm{h}},
\end{equation}
where $D_0$ is the diffusion coefficient of a single monomer of the
polymer chain, $N$ is the chain length, $\eta$ is the viscosity of the solvent,
and $R_\mathrm{h}$ is the hydrodynamic radius of the polymer. In general as
pointed out in Section \ref{sec:staticprops}, $R_\mathrm{h}$ is not linear in $N$,
resulting in a scaling different from $N^{-1}$ for the diffusion coefficient of
a polyelectrolyte in the presence of hydrodynamic interactions.


In Figure~\ref{fig:diffusion}, we compare the normalised diffusion
coefficient for polyelectrolyte chains of varying length with and
without hydrodynamic interactions.  When hydrodynamic interactions are
present, the polyelectrolyte diffusion shows a behaviour which can be
described by a power law scaling $D = D_0 N^{-m}$, where $D_0$ is the
monomer diffusion coefficient. For our model polyelectrolyte, a
scaling exponent of $m = 0.63 \pm 0.01$ is observed. Another way to
obtain the diffusion coefficient from is to insert the statically (or
dynamically) measured hydrodynamic radii and insert them into
Equation~\ref{eq:kirkwooddiffusion}. Both ways of obtaining the PE
diffusion constant are in perfect agreement, demonstrating the
applicability of Zimm theory. This is especially remarkable, as it
shows that the counterions, which are not considered in
Equation~\ref{eq:kirkwooddiffusion}, do not directly influence the
diffusion of polyelectrolytes. Instead only the conformations of the
polyelectrolyte itself, i.e. the internal monomer-monomer distances,
determine the diffusive behaviour.

The observed scaling exponent is in good agreement with experimental
results for the diffusion coefficient of polyelectrolytes as we have
shown earlier~\cite{grass08a}. For the standard polyelectrolyte PSS
(polystyrene sulfonate), B\"ohme and Scheler~\cite{boehme07b} obtained
$m = 0.64$, whereas Stellwagen et.~al.~\cite{stellwagen03a} reported a
scaling with $m = 0.617$. It is worth mentioning that the latter value
was obtained in the presence of 50 to 100 mM additional salt and for
PSS chains of upto 20000 repeat units, therefore a smaller scaling
exponent can be expected as shown in Section~\ref{sec:staticprops}.

When hydrodynamic interactions are switched off, the polyelectrolyte chain
exhibits Rouse behaviour with a scaling of $m = 1.02 \pm 0.02$, demonstrating
the importance of hydrodynamic interactions for the observed diffusion scaling
with chain length.

However, as shown in the previous chapter, the static chain
properties, including the hydrodynamic radius $R_\mathrm{h}$ do not
depend on the presence of hydrodynamic interactions. This allows for
the correct calculation of the diffusion coefficient even in the
absence of hydrodynamic interactions by means of
Equation~\ref{eq:kirkwooddiffusion}. Having said that, the direct
measurement of $D$ is only possible, when hydrodynamic interactions
are included in the simulation model.

For this study, we also lowered the monomer concentration of the
polyelectrolyte from $c_\mathrm{m} = 100$ mM up to 1 mM. Within this
range, no change of diffusion coefficients was observed (data not
shown).

\subsubsection{Influence of the electric field}


When an external electric field is applied to the polyelectrolyte solution
the diffusive motion in the direction of the electric field is mixed with the
induced directed motion. However, it is possible to determine the translational
diffusion coefficient $D_\mathrm{trans}$ perpendicular to the electric field.
Figure~\ref{fig:difftrans} shows that for a reduced electric field of $E =
0.1$ no deviation from the diffusive behaviour at vanishing electric field is
found. This is in-line with the findings of Section~\ref{sec:staticprops},
\ie that for small enough electric fields, the conformation of the polyelectrolyte
chains and the surrounding counterions are unchanged, such that the measured
quantities do not depend on the applied electric field.

Experimentally~\cite{boehme03a}, it is also possible to determine the diffusion
coefficient in the direction of the field by separating the diffusive from the
directed motion. The electric field showed no influence on the diffusion
coefficient.

\subsection{Electrophoretic mobility}\label{sec:electrophoreticmobility}

The second transport coefficient of interest is the electrophoretic mobility
$\mu$. It characterizes the motion of the polyelectrolyte in an external
electric field.

In capillary electrophoresis experiments~\cite{righetti96a}, the
electrophoretic mobility of the solute is determined by $\mu = \frac{v}{E} =
\frac{L l}{V t}$, where $v$ is the velocity, $E$ is the electric field, $V$ is
the applied voltage, $L$ is the total length of the capillary, $l$ is the
migration (or effective) length up to the detector, and $t$ is the detection
time of the solute.

This method can be directly transferred and applied to computer simulations. The
external electric field $E$ is modelled by a constant force acting on the
charged particles in the solution. This causes a directed motion with a certain
velocity $v$. From this one can obtain the electrophoretic mobility
\begin{equation}\label{eq:mobilitysimple}
	\mu = \frac{v}{E}.
\end{equation}

Netz~\cite{netz03b,netz03c} studied the behaviour of flexible polyelectrolytes
in strong electric fields in the absence of hydrodynamic interactions. He
showed, that the electrophoretic mobility determined by
Equation~\ref{eq:mobilitysimple} for large electric fields depends on the
magnitude of the electric field. This effect is attributed to the polarization
and the following removal of the counterion cloud surrounding the
polyelectrolyte. Below a critical value, the mobility is not affected by the
electric field, and the system is in the linear response regime.

Experimentally used electric fields are usually below 1 kV/cm, which
corresponds to a reduced field strength of $E = 0.001$, and as such are far
below this critical threshold. Thus we will focus on the behaviour of
polyelectrolytes in weak electric fields. In Section~\ref{sec:staticprops}, we
identified the threshold to be $E_\mathrm{crit} \approx 0.2$. For simulations,
the usage of a weak electrical driving force requires long simulations times,
in order to be able to accurately separate the directed motion from the
thermal fluctuations.

Alternatively, the electrophoretic mobility can be calculated from the
following Green-Kubo relation
\begin{equation}\label{eq:mobilitygreenkubo}
	\mu = \frac{1}{3 k_\mathrm{B} T} \sum_i q_i \int_0^\infty \left< \vec{v}_{i}(0)
	\cdot \vec{v}_{\mathrm{cm}}(\tau) \right> d\tau,
\end{equation}
where the summation is over all charged particles (monomers and counterions) in
the system, and $\vec{v}_{i}$ is their individual velocity and $q_i$ their
charge. Here, $\vec{v}_{\mathrm{cm}}$ is the velocity of the center of mass of the
polyelectrolyte. This approach has been successfully applied in simulations of
charged colloids~\cite{lobaskin07a,duenweg08a}.

While there are several
theories~\cite{barrat96a,muthukumar96a,volkel95b,mohanty99a} that have
been used to describe qualitatively the experimentally observed
electrophoretic behaviour of various PEs, there are still some open
questions to address.  Recent experiments on strongly charged flexible
PEs, such as polystyrene sulfonate (PSS) and single-stranded DNA
(ssDNA) of well defined length have shown a characteristic behaviour
for the short chain free-solution mobility
$\mu$~\cite{stellwagen02a,hoagland99a,cottet00b,stellwagen03a}: after
an initial increase of the mobility with increasing length, $\mu$
passes through a maximum, and then decreases towards a constant
mobility for long chains.

The increase of $\mu$ for short chains and the long-chain limit
constant $\mu$  can be explained within
the theoretical approaches, but the origins for the maximum for intermediate
chains have not been accounted for yet. In~\cite{grass08a}, we showed that the
experimentally observed behaviour can be simulated using a coarse-grained MD
model. We also showed that the maximum is due to the complex hydrodynamic
interactions between the polyelectrolyte and its counterions and the solute.


In Figure~\ref{fig:mobnohd}, we illustrate that the maximum in the
electrophoretic mobility can only be reproduced when hydrodynamic interactions
are properly accounted for. The neglect of hydrodynamic interactions leads to a
decreasing electrophoretic mobility for short chains. This observation
was also made in a recent publication by Frank and Winkler~\cite{frank08a}.

In addition to the measured mobilities, Figure~\ref{fig:mobnohd} includes
a prediction for the mobility without hydrodynamic interactions based
on a local
force balance: without hydrodynamic interactions, every particle of the
polyelectrolyte-counterion complex is subject to the same frictional force
$F_\mathrm{Solvent} = -\Gamma_0 v$ that counterbalances the electric driving force
$F_\mathrm{Field} = q_i E$. From this we can easily obtain the following expression
for the electrophoretic mobility in absence of hydrodynamic interactions, where
$\mu_0 = 1/\Gamma_0$ is the mobility of a single monomer, $N$ is the length
of the polyelectrolyte, and $N_\mathrm{CI}$ is the number of condensed
counterions that move with the polyelectrolyte:
\begin{equation}\label{eq:langmob}
	\frac{\mu}{\mu_0} = \frac{N-N_\mathrm{CI}}{N+N_\mathrm{CI}}.
\end{equation}

In order to utilize Equation~\ref{eq:langmob} we need the number of
condensed counterions. These  we obtain
$N_\mathrm{CI}$ by counting the average number of counter-ions found within $2
\sigma_0$ of the chain. Additionally, we can substitute $N_\mathrm{CI}$ from
Eq.~\ref{eq:manningnci} and obtain the Manning prediction for the mobility
\begin{equation}\label{eq:manningmob}
	\frac{\mu}{\mu_0} = \frac{1}{2\xi -1} \approx 0.2.
\end{equation}
Figure \ref{fig:mobnohd} shows excellent agreement of of
Equation~\ref{eq:langmob} and also demonstrates, that the Manning
mobility is approached nicely for long chains.

We conclude that the local force picture successfully describes the observed behaviour in
absence of hydrodynamic interactions, but qualitatively fails to describe the
mobility for shor chains in any real experiment, where hydrodynamic interactions
are obviously present.

\subsubsection{Influence of the electric field}


The effect of high electric fields on the mobility of polyelectrolytes has been
investigated theoretically before. Below the critical field strength, i.e.~in
the linear response regime, the electrophoretic mobility $\mu$ is independent
of the applied electric field. In Figure~\ref{fig:mobilityerror}, we compare
$\mu$ for a polyelectrolyte chain at $c_\mathrm{m}$ obtained at zero field via
Equation~\ref{eq:mobilitygreenkubo} to the mobility at finite fields, $E =
0.05$ and 0.1, via Equation~\ref{eq:mobilitysimple}. The measured values agree
within their displayed precision. For comparison, all data sets were obtained
using the same computational effort (same number of simulation steps). For
short chains, the Green-Kubo based method yields more accurate results, whereas
longer chains can be simulated at equal accuracy by an external applied field.
For short chains and weak fields, the thermal (Brownian) motion dominates the
directed electrophoretic motion, which decreases the accuracy of the obtained
values.

Based on this observation, we conclude that the Green-Kubo method to measure
the electrophoretic mobility of polyelectrolytes in solution has several
advantages. Firstly and most importantly, it guarantees the measurement of
free-solution mobilities in the linear regime, which are comparable to
experimental measurements at standard field strengths. The alternative method
uses fields that are about a factor 100 higher than the experimentally used ones and
are close to the critical value at which static and dynamic properties of the
polyelectrolyte are significantly changed. Secondly, the computational effort
needed to achieve a given accuracy for short polyelectrolytes chains ($N < 10$)
is up to 50\% smaller than with the direct method. And last but not least, the trajectories
simulated at zero field can be used to determine the electrophoretic mobility
and the diffusion (using Equation~\ref{eq:diffvac}) at the same time,
additionally reducing the computational cost of such a study.

\subsubsection{Influence of the monomer concentration}


It is well known, that the free-solution mobility of polyelectrolytes is
depending on the salt concentration of the solution~\cite{hoagland99a}. With
increasing additional salt, the maximal free-solution mobility decreases due
to increased counterion condensation. We point out that the long chain
mobility also shows a dependence on the monomer concentration in the
absence of additional salt. This effect can be mainly attributed to the
electrostatic screening mediated bu the chain monomers and counterions
(self-screening).

Figure~\ref{fig:mobilitycm} shows that not only the long chain
limit is influenced by increased electrostatic screening, but also the short
chain behaviour. The maximum is significantly reduced for higher monomer
concentrations. From a value higher or equal to $c_\mathrm{m} = 100$ mM it is completely reduced, and the
behaviour in the presence of hydrodynamic interactions is similar to one
without hydrodynamic interactions as seen in Figure~\ref{fig:mobnohd}. The
increased electrostatic screening caused by a higher concentration (resulting
in a shorter Debye length) suppresses the short-range hydrodynamic interactions
that are essential for the formation of the maximum.


This is reflected in Figure~\ref{fig:mobilitynmax}, where we show the degree
of polymerization with the maximum mobility $N_\mathrm{max}$ versus the Debye length
$\lambda_\mathrm{D}$. The lower the Debye length, i.e.~the higher the
electrostatic screening, the more the maximum is shifted to shorter chains. This
correlation has not been investigated previously. In the remainder of
this paper, we will analyze the importance of the interplay of the electrostatic
screening with the hydrodynamic interactions.

\section{Charge estimators}\label{sec:chargeestimators}

In this section, we present five different estimators to measure the
effective charge $Q_\mathrm{eff}$ of the polyelectrolyte counterion
complex. The first two are based on static ion distributions, whereas
the latter three are based on dynamic quantities that can be computed
with or without hydrodynamic interactions included. We will
compare the practicability and accuracy of the estimators and discuss
the obtained results in the context of polyelectrolyte
electrophoresis. At the same time we can see if any difference can be
observed between a static effective charge and a dynamic effective
charge~\cite{palberg04a,lobaskin07a}. 

\subsection{Primitive estimate}

\begin{figure}[ht]
\begin{center}
  \includegraphics[width=1.0\columnwidth]{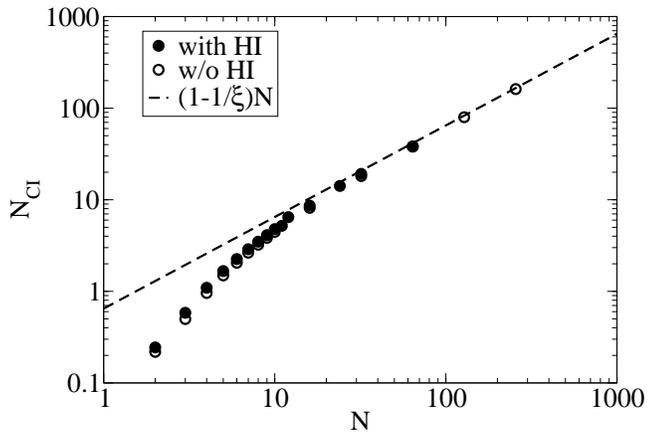}
  \caption{As a simple estimate of the number of condensed counterions
  $N_\mathrm{CI}$ all ions within a distance of $2 \sigma_0$ around the PE chain
  are summed up. For long polyelectrolyte chains, the value predicted by
  Equation~\ref{eq:manningnci} is approached. The static counterion
  distributions are not influenced by hydrodynamic interactions (HI).}
  \label{fig:nci}
\end{center}
\end{figure}

\begin{figure}[ht]
\begin{center}
  \includegraphics[width=1.0\columnwidth]{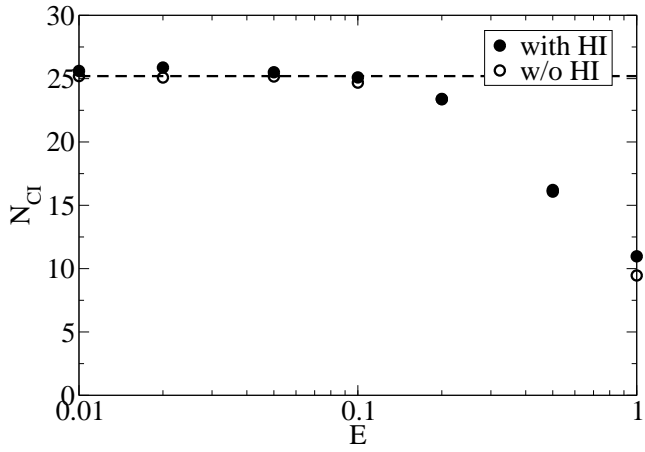}
  \caption{The number of
  condensed counterions $N_\mathrm{CI}$ for a polyelectrolyte chain of $N=40$
  depends on the applied external electrical field. Above a threshold of $E=0.1$, counterions start
  to be stripped from the PE chain and $N_\mathrm{CI}$ differs from the value at
  zero field (dashed line). This effect is independent of hydrodynamic
  interactions (HI).}
  \label{fig:nciE}
\end{center}
\end{figure}

\begin{figure}[ht]
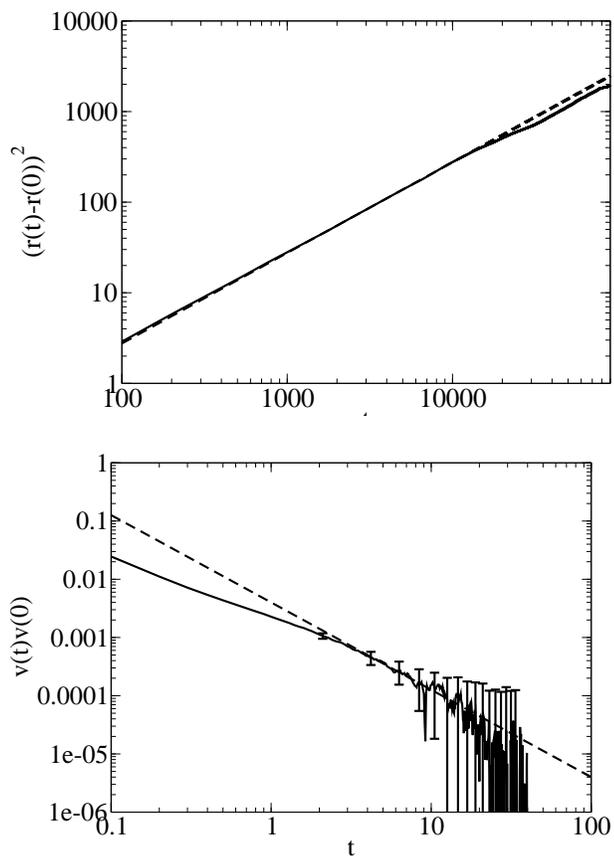

\begin{center}
  \subfigure{
    \includegraphics[width=.95\columnwidth]{diffusion-msdfit_080718}
    \label{fig:diffmethodsa}
  }
  \subfigure{
    \includegraphics[width=0.95\columnwidth]{diffusion-vacfit_080718}
    \label{fig:diffmethodsb}
  }
  \caption{The diffusion coefficient $D$ of a polyelectrolyte chain of length
  $N = 32$ is determined. (a) A fit (dashed line) to the linear part of the mean square
  displacement yields a diffusion
  coefficient of $D = 0.0045 \pm 0.0002$ via Equation~\ref{eq:diffmsd}. (b) By
  fitting a $t^{-3/2}$ power law to the long-time tail of the center of mass 
  auto-correlation function and using Equation~\ref{eq:diffvac} 
  we obtain $D = 0.0041 \pm 0.0005$.}
  \label{fig:diffmethods}
\end{center}
\end{figure}

\begin{figure}[ht]
\begin{center}
  \includegraphics[width=1.0\columnwidth]{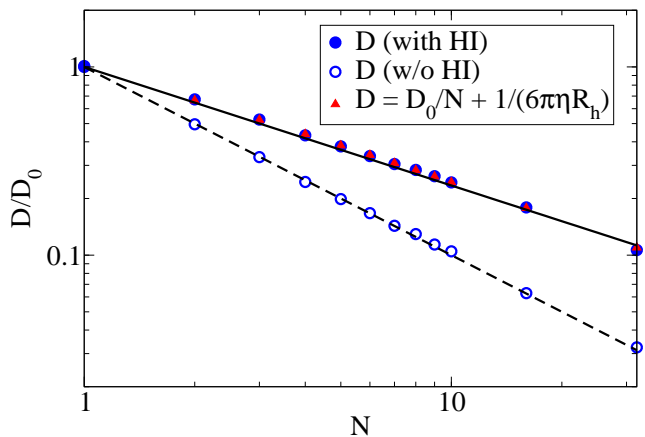}
  \caption{The diffusion of a polyelectrolyte chain of length
  $N$ at a monomer concentration $c_\mathrm{m} = 10$mM, normalised by the monomer
  diffusion $D_0$, shows the influence of hydrodynamic interactions (HI). With HI a scaling
  exponent of $m = 0.63 \pm 0.01$ is obtained (solid line), whereas without HI,
  the slope is $m = 1.02 \pm 0.02$ (dashed line). The diffusion in presence of 
 HI agrees with the values obtained from Equation~\ref{eq:kirkwooddiffusion},
 where $R_\mathrm{h}$ is determined from the simulation (triangles). }
  \label{fig:diffusion}
\end{center}
\end{figure}

\begin{figure}[ht]
\begin{center}
  \includegraphics[width=1.0\columnwidth]{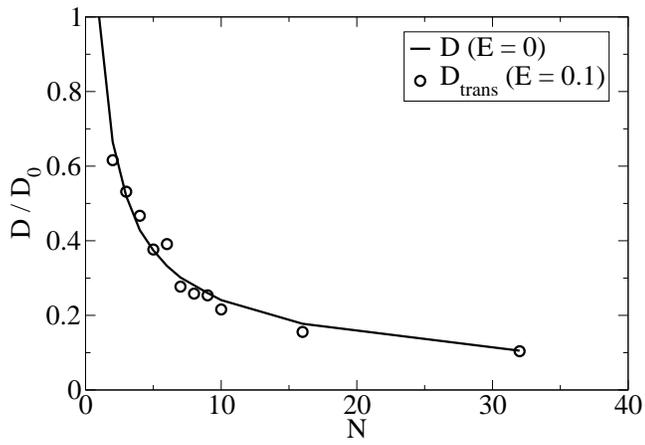}
  \caption[]{The translational diffusion coefficient measured at a reduced
  electric field $E=0.1$ (circles) is in agreement with the measurements at
  zero electric field (solid line).}
  \label{fig:difftrans}
\end{center}
\end{figure}

\begin{figure}[ht]
\begin{center}
  \includegraphics[width=1.0\columnwidth]{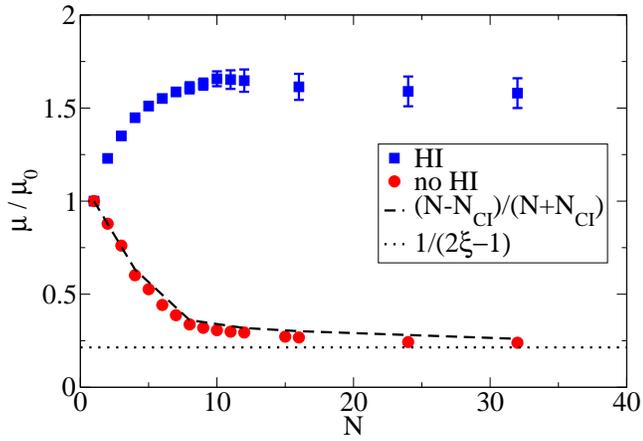}
  \caption[]{The normalised electrophoretic mobility at a monomer concentration
  of $c_\mathrm{m} = 5$ mM with hydrodynamic interactions (HI) reproduces the
  experimentally observed behaviour and shows a maximum for intermediate
  chains. Without HI the measures mobility strongly deviates from this showing
  a decrease with increasing chain length $N$. This behaviour can be explained
  by Equation \ref{eq:langmob} in the local force picture (dashed line) and
  approaches the limiting value of the Manning prediction
  (Eq.~\ref{eq:manningmob}). }
  \label{fig:mobnohd}
\end{center}
\end{figure}

\begin{figure}[ht]
\begin{center}
  \includegraphics[width=1.0\columnwidth]{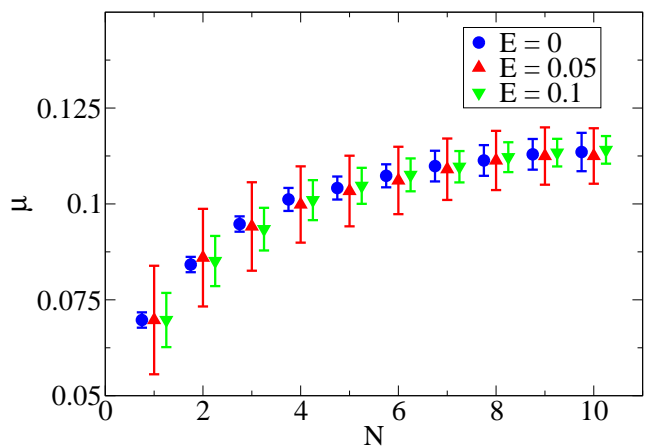}
  \caption[]{Electrophoretic mobility $\mu$ of polyelectrolyte chains at
  monomer concentration $c_\mathrm{m} = 5$ mM measured without electric field
  (circles) and with field at $E = 0.05$ and 0.1 (triangles) with the
  corresponding error bar. (Data sets have been shifted to increase
  readability.)}
  \label{fig:mobilityerror}
\end{center}
\end{figure}

\begin{figure}[ht]
\begin{center}
  \includegraphics[width=1.0\columnwidth]{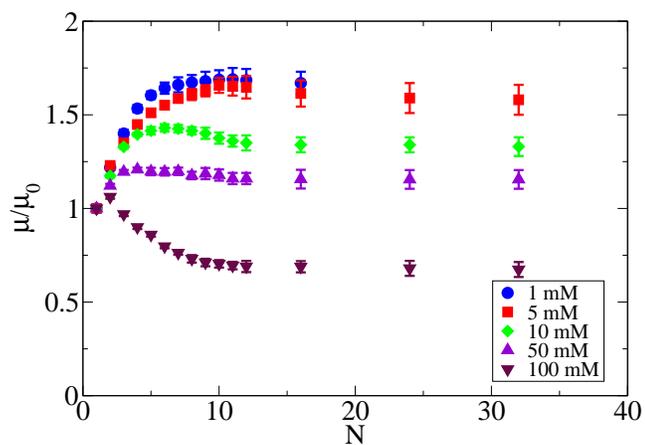}
  \caption[]{The normalised electrophoretic mobility $\mu/\mu_0$ shows a strong
  dependence on the monomer concentration $c_\mathrm{m}$. For dilute systems, the
  experimental behaviour is recovered, while at high monomer concentrations, the
  shape is significantly altered.}
  \label{fig:mobilitycm}
\end{center}
\end{figure}

\begin{figure}[ht]
\begin{center}
  \includegraphics[width=1.0\columnwidth]{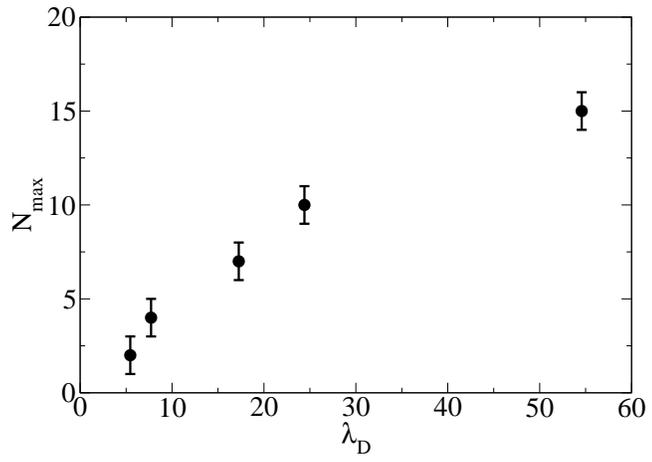}
  \caption[]{The approximate position of the maximum $N_\mathrm{max}$ changes with
  the Debye length $\lambda_\mathrm{D}$.}
  \label{fig:mobilitynmax}
\end{center}
\end{figure}

A simple method of estimating the effective charge has been used in
Figure~\ref{fig:nci} for describing the static properties of the
polyelectrolyte counterion complex:
\begin{equation}\label{eq:qeff1}
	Q_\mathrm{eff}^{(1)} = N_\mathrm{PE} - N_\mathrm{CI}(d<d_0),
\end{equation}
where $N_\mathrm{CI}(d<d_0)$ is the average number of counterions that can be
found within a distance $d$ around the polyelectrolyte, where $d$ is distance
to the closest monomer. The threshold $d_0$ is usually chosen to be $d_0
= 2 \sigma_0$.

In Section~\ref{sec:counterions}, we showed that the effective charge estimated in
this way is not affected by hydrodynamic interactions.

This method is computationally inexpensive and as seen in Figure~\ref{fig:nci}
yields a reasonable estimate for the effective charge. The draw back is that the
threshold $d_0$ is arbitrarily defined which limits the usefulness of this
estimator.

\subsection{Inflection criterion}

A more advanced method uses the inflection criterion to estimate the threshold of
counterion condensation~\cite{belloni84a,belloni98a,deserno00a} that
has its basis in Poisson-Boltzmann theory. It has been
shown that the position $d_\mathrm{c}$ of the inflection point (zero of the second
derivative) for the integrated ion distribution with respect to the logarithmic
distance from the closest chain monomer yields a cutoff that accurately
separates free from bound counterions in the case of a rod-like polyelectrolyte
in cylindrical geometry. The applicability to polyelectrolytes will be shown
here.


The integrated ion distribution is defined as
\begin{equation}
	I(d) = \int_0^d \rho_\mathrm{CI}(r) dr = f_\mathrm{CI}(r<d),
\end{equation}
where $\rho_\mathrm{CI}$ is the normalised density of counterions at a distance
$r$ to the closest chain monomer, and $f_\mathrm{CI}(r<d)$ is the fraction of
counterions found at a distance closer than $d$. The total number of condensed
counterions up to this distance given by multiplying this number with $N_\mathrm{CI}$.

\begin{figure}[ht]
\begin{center}  
  \includegraphics[width=1.0\columnwidth]{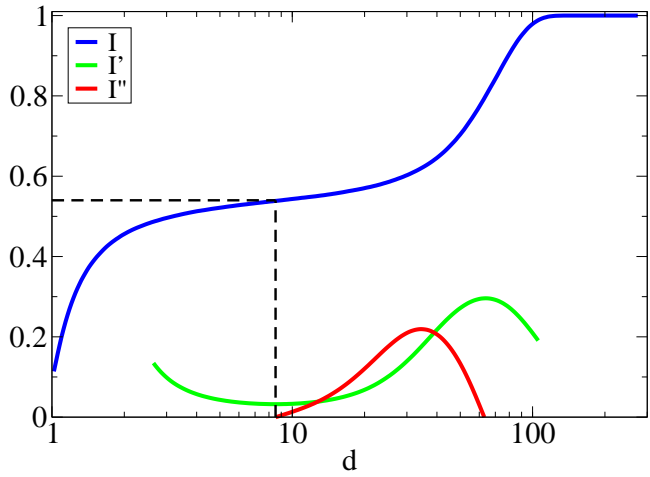}
  \caption{The integrated ion distribution $I(d)$ for a chain of $N=32$ monomers shows a clear inflection point identified by the zero of the second derivative with respect to $\ln(d)$. The value of $I(d)$ at the inflection point specifies the fraction of condensed counterions $f_\mathrm{CI}$.}
  \label{fig:effectivecharge-inflection}
\end{center}
\end{figure}

In order to apply the inflection point criterion to the polyelectrolyte model,
the integrated ion distribution has to be measured and plotted logarithmically.
Figure~\ref{fig:effectivecharge-inflection} shows the result for a chain of
$N=32$ monomers. Additionally, the first and second derivative obtained
numerically are plotted. The cutoff value $d_\mathrm{c}$ indicated by the inflection
point and the associated fraction of condensed counterions
$f_\mathrm{CI}$ can be directly read off Figure~\ref{fig:effectivecharge-inflection}.

The effective charge of the polyelectrolyte using the inflection criterion is
then given by
\begin{equation}\label{eq:qeff1b}
	Q_\mathrm{eff}^{(2)} = N_\mathrm{PE} - N_\mathrm{CI}(d<d_\mathrm{c}).
\end{equation}

The cutoff $d_\mathrm{c}$ is not a fixed parameter anymore, but has to be
determined for every chain length.

\subsection{Langevin model}

In Section~\ref{sec:electrophoreticmobility}, we used the local force picture
and derived Equation~\ref{eq:langmob} for the electrophoretic mobility in
absence of hydrodynamic interactions. Similarly, we can derive an expression
for the effective charge of the polyelectrolyte in absence of hydrodynamic
interactions based on the measured mobility $\mu$ of the PE-counterion
complex.

\begin{equation}\label{eq:qeff2}
	Q_\mathrm{eff}^{(3)} = N_\mathrm{PE}\left( 1 - \frac{1 - \mu \Gamma_0}{1 + \mu
	\Gamma_0}\right).
\end{equation}

Equation~\ref{eq:qeff2} offers a way to determine the effective charge
of the polyelectrolyte counterion complex based purely on the measured
mobility in the absence of hydrodynamic interactions without the need
of free parameter and at reasonable computational costs. The accuracy
of this estimator is limited by the accuracy of the measured
electrophoretic mobility $\mu$, and it is strictly valid only for the
Langevin model without HI.

\subsection{Ion diffusion}


The local force picture used for the Langevin model assumes that a finite
amount of counterions is bound to the polyelectrolyte while the remaining ions
can move freely. We will now motivate another charge estimator that is
especially promising as it can also be directly applied in experimental
setups, since all necessary quantities can be measured experimentally.

The $N_\mathrm{CI}$ bound counterions are expected to diffuse together with the
polyelectrolyte with a diffusion coefficient $D_\mathrm{PE}$ as determined in
Section~\ref{sec:diffusion}. Likewise, the $N-N_\mathrm{CI}$ free counterions will
diffuse with a different diffusion coefficient $D_0$. If one measures the ion
diffusion coefficient $D_\mathrm{CI}$ for all $N$ ions in such a system, the
measured quantity will be the weighted average of the bound and the free
diffusion coefficient:
\[
	D_\mathrm{CI} = \frac{N_\mathrm{CI} D_\mathrm{PE} + (N-N_\mathrm{CI}) D_0}{N}.
\]

From this we derive a novel estimator for the effective charge
\begin{equation}\label{eq:qeff3}
	Q_\mathrm{eff}^{(4)} = N \left( 1 - \frac{D_0-D_\mathrm{CI}}{D_0 -
	D_\mathrm{PE}} \right),
\end{equation} that has the advantage of only including quantities that are
experimentally accessible and as such can be used to directly measure the
effective charge in experiments.

\begin{figure}[ht]
\begin{center}
  \includegraphics[width=1.0\columnwidth]{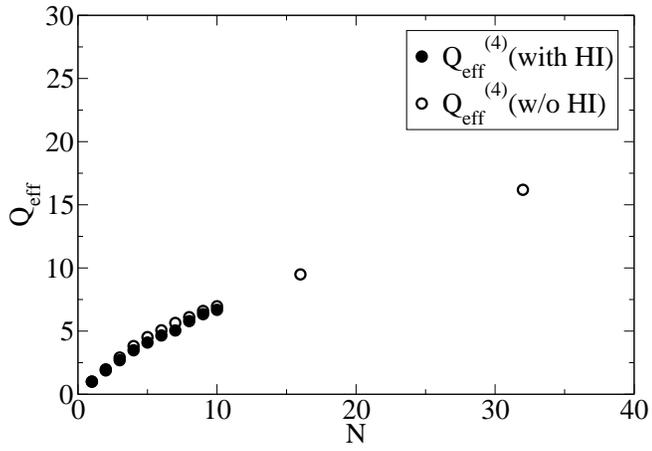}
  \caption[]{Estimated charge $Q_\mathrm{eff}^{(4)}$ of polyelectrolyte chains
  of varying length $N$ using Equation~\ref{eq:qeff3} with and without hydrodynamic
  interactions (HI).}
  \label{fig:qeff-diffestimator}
\end{center}
\end{figure}

In Figure~\ref{fig:qeff-diffestimator} we compare the results of estimator
$Q_\mathrm{eff}^{(4)}$ with and without hydrodynamic interactions and that
within the accuracy of the method, no difference can be observed, even though the
individual diffusion coefficients are different for both types of simulations.
Equation~\ref{eq:qeff3} is valid for simulations independent of the presence
or absence of hydrodynamic interactions and has the computational complexity of
determining the diffusion coefficients (see Section~\ref{sec:diffusion}).

This estimator is especially interesting, as it only involves quantities that
are directly obtainable from experiments. As such, it might provide a
great opportunity to measure the effective charge of macromolecules.
To our knowledge, this method has not been suggested before. The derivation of
equation \ref{eq:qeff3} assumes the absence of additional salt, i.e. $c_s = 0$,
but can be modified to account for a finite amount of additional salt. A
similar analysis can be done using the free and the bound mobility, but these
quantities are experimentally more difficult to obtain.

\subsection{Co-moving counterions}


The last estimator for the effective charge we introduce in this article is
based on directly determining the counterions that are co-moving with the
polyelectrolyte during free-solution electrophoresis similar to the method used
in~\cite{lobaskin04a,chatterji07a}.

When applying an external electric field $E$, the polyelectrolyte moves with a
velocity $v_\mathrm{PE} = \mu_\mathrm{PE} E$ in the direction of the electric field.
Co-moving counterions move with the same velocity in the same direction, whereas free
counterions move with a velocity $v_\mathrm{CI,0} = \mu_0 E$ into the opposite
direction. Here, $\mu_0$ is the mobility of free counterions.

\begin{figure}[ht]
\begin{center}
  \includegraphics[width=1.0\columnwidth]{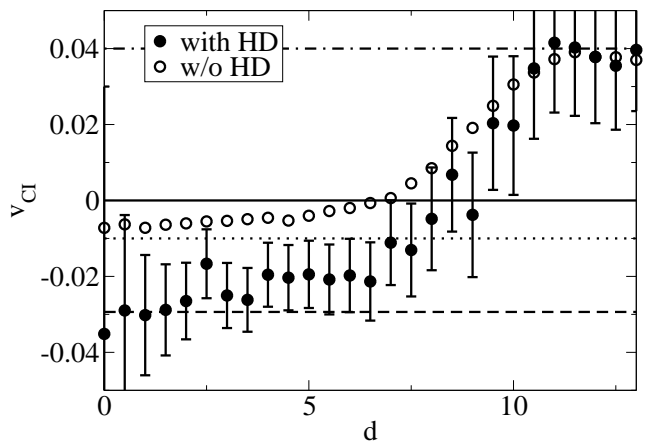}
  \caption[]{The ion velocity distribution $v_\mathrm{CI}(d)$ for a
  polyelectrolyte chain of length $N=32$ shows the transition from co-moving
  counterions with a velocity similar to $v_\mathrm{PE}$ (dashed line and dotted
  line indicate $v_\mathrm{PE}$ with and without hydrodynamic interactions (HI))
  to the free ion velocity $v_\mathrm{CI,0}$ (dash-dot line). $v_\mathrm{CI}(d=d_0)
  = 0$ defines the threshold between co- and counter moving ions and is used to
  determine the effective charge of the polyelectrolyte counterion complex in
  Equation~\ref{eq:qeff4}.}
  \label{fig:ionvelocity}
\end{center}
\end{figure}

In Figure~\ref{fig:ionvelocity}, we plot the velocity $v_\mathrm{CI}$ of the ions
versus the distance $d$ to the center of mass of the polyelectrolyte. It can be
seen, that ions close to the center of mass are co-moving, i.e. $v_\mathrm{CI} =
v_\mathrm{PE}$, and ions far away from the chain are indeed freely moving with a
velocity $v_\mathrm{CI} = v_\mathrm{CI,0}$. For an intermediate value of $d_0$,
the $v_\mathrm{CI} = 0$, i.e. the ions are not moving relative to the electric
due to the interaction with the polyelectrolyte. This distance $d_0$ separates
the co-moving counterions from the free-moving ones. Similar to
$Q_\mathrm{eff}^{(1)}$ (Eq.~\ref{eq:qeff1}), the effective charge is obtained by
averaging the number of counterions found within a distance $d_0$ of the center
of mass of the chain.
\begin{equation}\label{eq:qeff4}
	Q_\mathrm{eff}^{(5)} = N_\mathrm{PE} - N_\mathrm{CI}(d<d_0),
\end{equation}
Note, that for this estimator $d$ is the distance of the counterion to the
center of mass of the polyelectrolyte, not to the closest polyelectrolyte
monomer.

The results shown in Figure~\ref{fig:ionvelocity} indicate a small influence of
the hydrodynamic interactions on the value of $d_0$, but it is important to note that the ion
density in this region is very low resulting in almost identical values of the
$Q_\mathrm{eff}^{(5)}$ in both cases.

The threshold used for this estimator is not predefined, but has to be
determined from the simulation and generally is a function of the chain length.
Since $v_\mathrm{CI}$ has to be determined for each distance $d$ to the center of
mass of the polyelectrolyte requiring high statistics.

\subsection{Comparison}

\begin{figure}[ht]
\begin{center}
  \includegraphics[width=1.0\columnwidth]{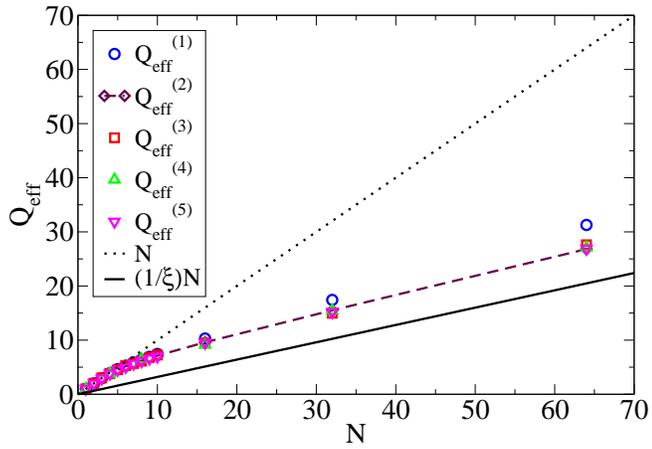}
  \caption[]{The effective charge of polyelectrolyte chains of length $N$ using the
  estimators $Q_\mathrm{eff}^{(1)}$ to $Q_\mathrm{eff}^{(5)}$. The dotted line
  indicates the bare, unscreened charge of the polyelectrolyte, whereas the solid
  line shows a prediction based on counterion condensation theory,
  with $\xi = 2.84$ being the condensation parameter. Whereas estimators
  $Q_\mathrm{eff}^{(3)}$ to $Q_\mathrm{eff}^{(5)}$ agree over the range of
  lengths $N$ tested, $Q_\mathrm{eff}^{(1)}$ overestimates the effective charge
  of the polyelectrolyte counterion complex.}
  \label{fig:estimatorcomparison}
 \end{center}
\end{figure}

Figure~\ref{fig:estimatorcomparison} compares the results of all
presented charge estimators $Q_\mathrm{eff}^{(1)}$ to $Q_\mathrm{eff}^{(5)}$. For
short chains ($N<4$), the estimators agree and coincide with the bare,
unscreened charge of the polyelectrolyte. In this regime, no counterion condensation is observed.

For intermediate and long chains ($N>4$), the effective charge is reduced as it
deviates from the bare charge and tends towards the Manning
prediction. However, it probably will never reach it
completely as the polyelectrolyte starts to assume a coiled conformation that is
not accounted for in condensation theory. Here, the condensation parameter for
the polyelectrolyte system is $\xi = 2.84$. In this regime, the simple estimator
$Q_\mathrm{eff}^{(1)}$ measures a higher effective charge, \ie not all condensed
counterions that are correctly included in the other estimators are taken into
account. This effect is not observed if the cutoff value is chosen using the
inflection criterion.

All estimators show no or little influence of the hydrodynamic interactions on
the effective charge of the polyelectrolyte counterion complex. This
independence on the hydrodynamic interactions has been recently observed for
highly charged colloid using a similar simulation approach \cite{chatterji07a}.

Furthermore, no difference between the static estimate,
$Q_\mathrm{eff}^{(2)}$ and the dynamic estimates
$Q_\mathrm{eff}^{(3-5)}$ for the effect charge are observed. This is
especially remarkable as it is an open question if there is a
difference between the static and the dynamic effective charge. For
the case of charged colloids this seems to be the
case~\cite{palberg04a,lobaskin07a}, but the results of this work show
that for strongly charged linear polyelectrolytes both quantities are
identical and one does not have to differentiate between a statically
and dynamically renormalized charge.

Two of these estimators are particularly promising. The charge
estimator based on the Langevin model, $Q_\mathrm{eff}^{(3)}$, has the
advantage that it is only necessary to determine one dynamic quantity,
namely the electrophoretic mobility. All other parameters are directly
given by the model. In this way one can obtain an accurate estimate
without much computational effort. The ion-diffusion estimator
$Q_\mathrm{eff}^{(4)}$, on the other hand, requires the calculation of
three dynamic quantities, which is slightly more costly in simulation
but has the great advantage of being directly transferable to
experiments, since all needed quantities, \ie the free ion diffusion,
the bound ion diffusion, and the chain diffusion can be measured by
standard techniques. To the author's knowledge, this is the first time
this method is proposed to determine the effective charge of
polyelectrolytes during electrophoresis.

\section{Effective friction}\label{sec:effectivefriction}


We will now use the effective charge obtained in the previous section to quantify the effective
friction of the polyelectrolyte-counterion complex. As shown in
Figure~\ref{fig:estimatorcomparison}, the estimators $Q_\mathrm{eff}^{(2)}$ to
$Q_\mathrm{eff}^{(5)}$ yield the same effective charge, we thus only employ
$Q_\mathrm{eff}^{(3)}$ to determine the effective friction in the following
equation:
\begin{equation}\label{eq:effectivefriction}
	\Gamma_\mathrm{eff} = \frac{Q_\mathrm{eff}}{\mu}.
\end{equation}
Here, we use $\mu$ as obtained in Section~\ref{sec:electrophoreticmobility}.

\begin{figure}[ht]
\begin{center} 
  \includegraphics[width=1.0\columnwidth]{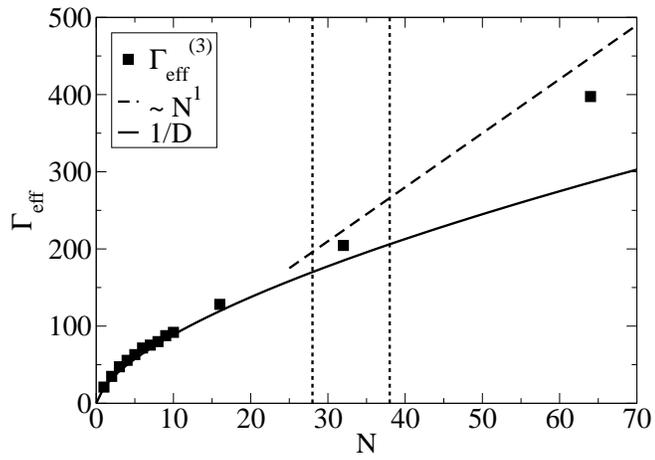}
  \caption[Effective friction]{The effective friction of the
  polyelectrolyte counterion complex is in agreement with the
  hydrodynamic friction ($\Gamma = 1/D$) for short chains only. For
  longer chains beyond the Debye length in the system (indicated by
  the dashed lines) the effective friction deviates and seems to
  approach a linear behaviour.}
  \label{fig:effective-friction}
\end{center}
\end{figure}

From the result displayed in Figure~\ref{fig:effective-friction} we
can see that for short chains the effective friction determined by
Equation~\ref{eq:effectivefriction} is in agreement with the
hydrodynamic polymer friction $\Gamma = 1/D$ that follows the Einstein
relation. For longer chains, on the other hand, a significant
deviation is observed. The effective friction no longer follows the
$N^{-m}$ behaviour of the Einstein relation (with $m\approx0.63$), but
instead tends towards a linear increase in N. The length scale
separating both regimes is of the order of the Debye length in the
system.

From this we infer that the increased effective friction is caused by
the counterions close to the polyelectrolyte chain, which destroy
long-range hydrodynamic interactions between distant parts of the
polyelectrolyte. In this way they also increase the hydrodynamic screening, and effectively decouple
the chain monomers from each other. An similar conclusion is obtained in an independent study by
Fischer et.~al.~\cite{fischer08a}. The effect is strongly related to
the counterion density in the vicinity of the chain and thus depends
on the Debye length in the system.

\begin{figure}[ht]
\begin{center}
  \includegraphics[width=1.0\columnwidth]{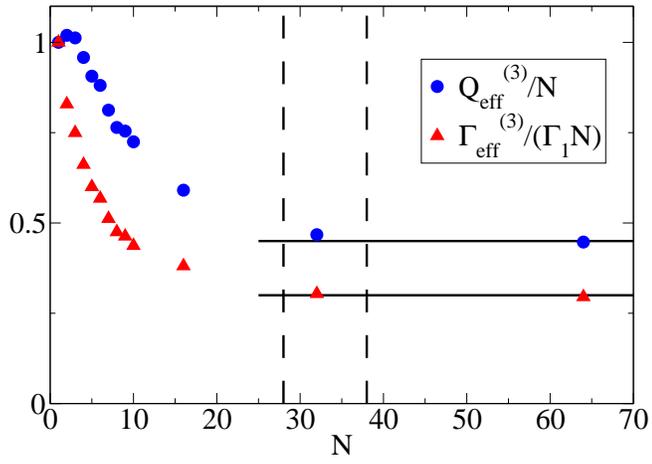}
  \caption[]{The effective charge $Q_\mathrm{eff}$ and the effective friction
  $\Gamma_\mathrm{eff}$ per monomer show a strong dependence on the chain length
  $N$ for short chains. For long chains, both values are constant. The
  transition occurs on a length-scale similar to the Debye length
  (dashed lines).}
  \label{fig:effectivecharge-monomer}
\end{center}
\end{figure}

It becomes even more evident when looking at the effective friction
per monomer, Figure~\ref{fig:effectivecharge-monomer}. Initially, for
short chains, the effective friction decreases strongly with
increasing chain length. The monomers move together through the
solvent and can shield each other from the flow and in this way
reducing the effective friction with the solvent. For longer chains,
this reduction of the effective friction becomes less effective
because the counterions that associate with the polyelectrolyte
influence the solvent flow around the polyelectrolyte effectively
canceling the beneficial shielding effects. On a length scale that is
comparable to the Debye length in the system, different parts of the
polyelectrolyte become hydrodynamically decoupled and the effective
friction per monomer does not depend on the length of the
polyelectrolyte any more. 

The effective charge per monomer shows a different behaviour. Short chains do
not have any bound counterions and show their bare charge of 1e per monomer.
With increasing length, they can attract counterions which are then co-moving with the
polyelectrolyte and reducing its effective charge. Again the concentration of
counterions around the polyelectrolyte plays an important role and the
relevant length-scale can be compared to the Debye length. The similarities in length scales of the
Debye length to the hydrodynamic screening length have been treated
analytically previously in Refs. \cite{long96a,viovy00a,tanaka02a}.

\begin{figure}[ht]
\begin{center}
  \includegraphics[width=\columnwidth]{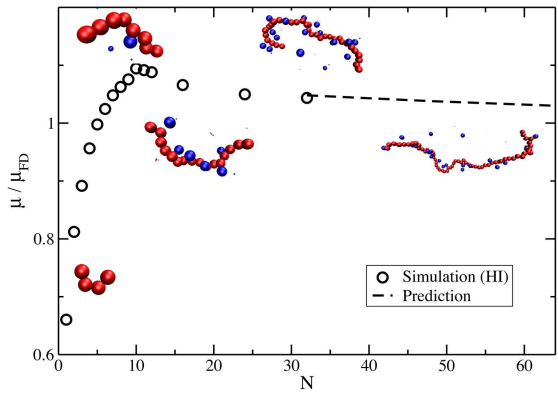}
  \caption{The interplay between counterion condensation and
  conformational change of the polyelectrolyte results in the combined behaviour
  of effective friction and effective charge that leads to the observed length-
 dependence of the electrophoretic mobility.}
  \label{fig:mobconform}
\end{center}
\end{figure}

The combined behaviour of effective friction and effective charge leads to the
observed length dependent mobility for short chains and the constant mobility
for long chains is illustrated in Figure~\ref{fig:mobconform}. Initially, very
few counterions are attracted to the polyelectrolyte and the rise in mobility
is due to the hydrodynamic interactions between the chain monomers, which are
in an extended conformation. Then, at intermediate chain lengths, the
counterion condensation increases, reducing the effective charge of the
polyelectrolyte, which leads to the observed maximum. For longer chains, the
counterions shield the electrostatic and hydrodynamic interactions between the
chain monomers and thereby induce a transistion to less extended conformations,
that for even longer chains (not shown here) become globular and coil-like. The
screening of hydrodynamic interactions together with the conformational change
increases the effective friction of the polyelectrolyte, leading to the
observed decrease in the electrophoretic mobility and the long-chain plateau
value.

More specifically, the mobility maximum, which is observed for flexible
polyelectrolytes, is due to the efficient shielding between monomers for short
chains which reduces the effective friction. Stiffer polyelectrolytes experience
a stronger friction since the monomers in rod-like conformations can not shield
each other that efficiently. This is the reason, why the maximum in the mobility
for intermediate chains is only observable for flexible or semi-flexible
polyelectrolytes such as PSS or single-stranded DNA but not for the more rigid
double-stranded DNA. On the other hand, the decrease of the effective charge by
counterion attraction is depending on the linear charge density of the
polyelectrolyte. The higher the polyelectrolyte is charge, the more the effective
charge is reduced by co-moving counterions. Additionally, we showed that the
counterions also increase the effective friction. Both effects work in the same
direction and cause the maximum to be shifted to shorter chains (in case of PSS)
or disappear completely (for double-stranded DNA).

\section{Conclusion}

We studied in detail the electrophoretic behaviour of flexible
polyelectrolyte chains by means of a coarse-grained molecular dynamics
model. The static chain properties have been investigated for various
values of the external electric field, and for different salt
concentrations. The static properties do not depend on the presence or
absence of hydrodynamic interactions. Our results show the expected
behaviour of polyelectrolyte chains: the scaling of $R_\mathrm{e,g,h}$
for short chains exhibits an effective exponent that lies between the
infinite dilution exponent, $\nu = 1$, and the Flory exponent for
uncharged chains, $\nu = 0.588$, and $\nu$ decreases with increasing
salt concentration due to electrostatic screening.

We tested the influence of the strength of an applied external
electric field on the chain conformations and the counterion
cloud. Below a critical value $E_\mathrm{crit}$, no dependence on the
field strength is observed, which is in agreement with linear response
theory. An applied electric field larger than a critical value of
$E_\mathrm{crit} \approx 0.2$ leads to extended chain conformations
and to a significant loss of condensed counterions. Our results agree
with the theoretical predictions by R.~R.~Netz~\cite{netz03b,netz03c},
which, for our system, yielded a value of $E_\mathrm{c} = 0.25$.

When simulating the dynamic transport properties, such as diffusion
coefficient and electrophoretic mobility, long-range hydrodynamic
interactions become crucial. Only when they are included, the
experimentally observed behaviour is reproduced and excellent
agreement between experiments and simulations is found for our PSS
model system. Our results demonstrate convincingly that it is possible
to model quantitavively the dynamic behaviour of polyelectrolytes
using coarse grained models.

To investigate the dynamics even further, we present five different
approaches to estimate the dynamical effective charge of the
polyelectrolyte-counterion complex during free-solution
electrophoresis. The charges calculated by the estimators are not
influenced by hydrodynamic interactions, nor by the fact if they were
obtained from static or dynamic observables. Thus we find no
difference between a static and a dynamic effective charge. In
addition, it is possible to use the estimators in simulations without
hydrodynamic interactions that are computationally inexpensive.

For the charge estimate based on the Langevin model,
$Q_\mathrm{eff}^{(2)}$, it is only necessary to determine one
dynamical quantity, the electrophoretic mobility. All other parameter
are directly given by the model. In this way one can obtain an
accurate estimate without too much computational effort. The
ion-diffusion estimator $Q_\mathrm{eff}^{(2)}$, on the other hand,
requires the calculation of three dynamical quantities, which is
slightly more costly in simulation but has the great advantage of
being directly transferable to experiments since all needed
quantities, i.e.~the free ion diffusion, the bound ion diffusion and
the chain diffusion can be measured by standard techniques. To our
knowledge, this is the first time, this method is proposed to
determine the effective charge of polyelectrolytes during
electrophoresis.

Using theses estimators, we determined the length dependence of the
effective charge, and combined it with the measurements of the
electrophoretic mobility to obtain the effective friction of the
polyelectrolyte. The results indicate that this effective friction
during electrophoresis is different from the hydrodynamic friction for
a single polyelectrolyte chain obtained from diffusion
measurements. We attribute this difference to the contribution by the
co-moving counterions, which cause a shielding of the hydrodynamic
interactions.  We identified a hydrodynamic screening length beyond
which the effective friction approaches a constant value per monomer
which - together with the constant value per monomer of the effective
charge - leads to the well-known and observed constant electrophoretic
mobility for long polyelectrolyte chains.  We showed that this
hydrodynamic screening range is comparable to the Debye length for
electrostatic screening with the system.

The results of this study provide an in-depth understanding of the
microscopic processes that govern the macroscopic behaviour of charged
polyelectrolytes in free solution. The characterisation of the
dynamical effective friction of macromolecules based on the direct
measurement of the effective charge seems to be a promising tool to
investigate and understand related systems, such as the dynamics of
polyelectrolytes slowed down by additional hydrodynamic drag tags
(ELFSE). A study of this topic is currently under preparation.

\section*{Acknowledgements}

We acknowledge inspiring discussions with U.~Scheler on the matter of charge
estimators. Furthermore, we thank B.~D\"{u}nweg, U.~Schiller, and G.~Slater
for helpful remarks.  Funds from the the Volkswagen foundation, the DAAD,
and DFG under the TR6 are gratefully acknowledged. All simulations were
carried out on the compute cluster of the Center for Scientic computing
at Goethe University Frankfurt.

\clearpage 

\clearpage
\begin{list}{}{\leftmargin 2cm \labelwidth 1.5cm \labelsep 0.5cm}

\item[\bf Fig. 1] Both the end-to-end distance
  $R_\mathrm{e}$ and the radius of gyration $R_\mathrm{g}$ at vanishing external field
  for a monomer concentration of
  $c_\mathrm{PE} = c_\mathrm{CI} = 10$ mM exhibit an effective scaling exponent $\nu =
  0.85$. The hydrodynamic radius $R_\mathrm{h}$ shows a different behaviour for short chains
  and only slowly reaches the asymptotic scaling.
  The static chain properties are not influenced by hydrodynamic
  interactions (HI) and fully agree with the ones obtained without HI.
\item[\bf Fig. 2] (a) The end-to-end distance $R_\mathrm{e}$ and (b) the radius
of gyration radius $R_\mathrm{g}$ for a chain of $N=40$ monomers differ from the
  value at zero external field $E$ (dashed line), if a critical
  value is reached. Beyond this threshold, the polyelectrolyte chain is extended
  (increased $R_\mathrm{e}$ and $R_\mathrm{g}$) and starts to align with the field
  directed in x-direction. This effect is independent of hydrodynamic interactions (HI).
\item[\bf Fig. 3] The end-to-end distance
  $R_\mathrm{e}$ and the radius of gyration $R_\mathrm{g}$ at zero external
  field and for a monomer concentration $c_\mathrm{PE} = 10$ mM and no added salt
  ($c_\mathrm{s} = 0$ M) exhibit an effective scaling exponent $\nu = 0.85$ (dashed
  line). With $c_\mathrm{s} = 1$ M monovalent salt added, the scaling exponent drops down to
  $\nu = 0.68$ (dotted line).
\item[\bf Fig. 4] The effective scaling factor $\nu$ shows strong dependence of
the Debye length. With out added salt (i.e.~large Debye lengths) $\nu$ is close to 1, whereas
  for high salt concentrations (i.e.~small Debye lengths) approaches the
  Flory number, the value of an uncharged polymer.
  \item[\bf Fig. 5] As a simple estimate of the number of condensed counterions
  $N_\mathrm{CI}$ all ions within a distance of $2 \sigma_0$ around the PE chain
  are summed up. For long polyelectrolyte chains, the value predicted by
  Equation~\ref{eq:manningnci} is approached. The static counterion
  distributions are not influenced by hydrodynamic interactions (HI).
  \item[\bf Fig. 6] The number of
  condensed counterions $N_\mathrm{CI}$ for a polyelectrolyte chain of $N=40$
  depends on the applied external electrical field. Above a threshold of $E=0.1$, counterions start
  to be stripped from the PE chain and $N_\mathrm{CI}$ differs from the value at
  zero field (dashed line). This effect is independent of hydrodynamic
  interactions (HI).
  \item[\bf Fig. 7] The diffusion coefficient $D$ of a polyelectrolyte chain of
  length $N = 32$ is determined. (a) A fit (dashed line) to the linear part of the mean square
  displacement yields a diffusion
  coefficient of $D = 0.0045 \pm 0.0002$ via Equation~\ref{eq:diffmsd}. (b) By
  fitting a $t^{-3/2}$ power law to the long-time tail of the center of mass 
  auto-correlation function and using Equation~\ref{eq:diffvac} 
  we obtain $D = 0.0041 \pm 0.0005$.
  \item[\bf Fig. 8] The diffusion of a polyelectrolyte chain of length
  $N$ at a monomer concentration $c_\mathrm{m} = 10$mM, normalised by the monomer
  diffusion $D_0$, shows the influence of hydrodynamic interactions (HI). With HI a scaling
  exponent of $m = 0.63 \pm 0.01$ is obtained (solid line), whereas without HI,
  the slope is $m = 1.02 \pm 0.02$ (dashed line). The diffusion in presence of 
 HI agrees with the values obtained from Equation~\ref{eq:kirkwooddiffusion},
 where $R_\mathrm{h}$ is determined from the simulation (triangles). 
 \item[\bf Fig. 9] The translational diffusion coefficient measured at a reduced
  electric field $E=0.1$ (circles) is in agreement with the measurements at
  zero electric field (solid line)
  \item[\bf Fig. 10] The normalised electrophoretic mobility at a monomer
  concentration of $c_\mathrm{m} = 5$ mM with hydrodynamic interactions (HI) reproduces the
  experimentally observed behaviour and shows a maximum for intermediate
  chains. Without HI the measures mobility strongly deviates from this showing
  a decrease with increasing chain length $N$. This behaviour can be explained
  by Equation \ref{eq:langmob} in the local force picture (dashed line) and
  approaches the limiting value of the Manning prediction
  (Eq.~\ref{eq:manningmob}).
  \item[\bf Fig. 11] Electrophoretic mobility $\mu$ of
  polyelectrolyte chains at monomer concentration $c_\mathrm{m} = 5$ mM measured without electric field
  (circles) and with field at $E = 0.05$ and 0.1 (triangles) with the
  corresponding error bar. (Data sets have been shifted to increase
  readability.)
  \item[\bf Fig. 12] The normalised electrophoretic mobility $\mu/\mu_0$ shows a
  strong dependence on the monomer concentration $c_\mathrm{m}$. For dilute systems, the
  experimental behaviour is recovered, while at high monomer concentrations, the
  shape is significantly altered.
  \item[\bf Fig. 13] The approximate position of the maximum $N_\mathrm{max}$
  changes with the Debye length $\lambda_\mathrm{D}$.
  \item[\bf Fig. 14] The integrated ion distribution $I(d)$ for a chain of
  $N=32$ monomers shows a clear inflection point identified by the zero of the
  second derivative with respect to $\ln(d)$. The value of $I(d)$ at the
  inflection point specifies the fraction of condensed counterions
  $f_\mathrm{CI}$.
  \item[\bf Fig. 15] Estimated charge $Q_\mathrm{eff}^{(4)}$ of polyelectrolyte
  chains of varying length $N$ using Equation~\ref{eq:qeff3} with and without
  hydrodynamic interactions (HI).
  \item[\bf Fig. 16] The ion velocity distribution $v_\mathrm{CI}(d)$ for a
  polyelectrolyte chain of length $N=32$ shows the transition from co-moving
  counterions with a velocity similar to $v_\mathrm{PE}$ (dashed line and dotted
  line indicate $v_\mathrm{PE}$ with and without hydrodynamic interactions (HI))
  to the free ion velocity $v_\mathrm{CI,0}$ (dash-dot line). $v_\mathrm{CI}(d=d_0)
  = 0$ defines the threshold between co- and counter moving ions and is used to
  determine the effective charge of the polyelectrolyte counterion complex in
  Equation~\ref{eq:qeff4}.
  \item[\bf Fig. 17] The effective charge of polyelectrolyte chains of length
  $N$ using the estimators $Q_\mathrm{eff}^{(1)}$ to $Q_\mathrm{eff}^{(5)}$. The dotted line
  indicates the bare, unscreened charge of the polyelectrolyte, whereas the solid
  line shows a prediction based on counterion condensation theory,
  with $\xi = 2.84$ being the condensation parameter. Whereas estimators
  $Q_\mathrm{eff}^{(3)}$ to $Q_\mathrm{eff}^{(5)}$ agree over the range of
  lengths $N$ tested, $Q_\mathrm{eff}^{(1)}$ overestimates the effective charge
  of the polyelectrolyte counterion complex.
  \item[\bf Fig. 18] The effective friction of the
  polyelectrolyte counterion complex is in agreement with the
  hydrodynamic friction ($\Gamma = 1/D$) for short chains only. For
  longer chains beyond the Debye length in the system (indicated by
  the dashed lines) the effective friction deviates and seems to
  approach a linear behaviour.
  \item[\bf Fig. 19] The effective charge $Q_\mathrm{eff}$ and the effective
  friction $\Gamma_\mathrm{eff}$ per monomer show a strong dependence on the chain length
  $N$ for short chains. For long chains, both values are constant. The
  transition occurs on a length-scale similar to the Debye length
  (dashed lines).
  \item[\bf Fig. 20] The interplay between counterion condensation and
  conformational change of the polyelectrolyte results in the combined behaviour
  of effective friction and effective charge that leads to the observed length-
  dependence of the electrophoretic mobility.
\end{list}

\clearpage


\clearpage

\clearpage

\clearpage

\clearpage

\clearpage

\clearpage

\clearpage

\clearpage

\clearpage

\clearpage

\clearpage

\clearpage

\clearpage

\clearpage

\clearpage

\clearpage

\clearpage

\clearpage

\clearpage 

\end{document}